\documentclass[twocolumn,superscriptaddress,showpacs,preprintnumbers,amsmath,amssymb]{revtex4}
\usepackage{graphicx}
\usepackage{dcolumn}
\usepackage{bm}
\usepackage{epsf}
\usepackage{dcolumn}
\def\be{\begin{equation}}
\def\ee{\end{equation}}
\def\bea{\begin{eqnarray}}
\def\eea{\end{eqnarray}}
\def\gsim{\ \rlap{\raise 2pt\hbox{$>$}}{\lower 2pt \hbox{$\sim$}}\ }
\def\lsim{\ \rlap{\raise 2pt\hbox{$<$}}{\lower 2pt \hbox{$\sim$}}\ }
\def\dslash{\kern-4pt \not{\hbox{\kern-2pt $\partial$}}}
\def\pslash{\not{\hbox{\kern-2pt p}}}

\begin{document}
\DeclareGraphicsExtensions{.eps,.ps}


\title{Neutrino mass hierarchy and octant determination with atmospheric neutrinos}



\author{Vernon Barger}
\affiliation{Department of Physics, University of Wisconsin-Madison,
1150 University Avenue Madison, Wisconsin 53706, USA}

\author{Raj Gandhi}
\affiliation{
Harish-Chandra Research Institute, Chhatnag Road, Jhunsi,
Allahabad 211 019, India}

\author{Pomita Ghoshal}
\affiliation{
Harish-Chandra Research Institute, Chhatnag Road, Jhunsi,
Allahabad 211 019, India}
 
\author{Srubabati Goswami}
\affiliation{
Physical Research Laboratory, Navrangpura,
Ahmedabad 380 009, India}

\author{Danny Marfatia}
\affiliation{Department of Physics and Astronomy, University of Kansas,
1082 Malott Hall, 1251 Wescoe Hall Drive, 
Lawrence, Kansas 66045, USA}

\author{Suprabh Prakash}
\affiliation{
Department of Physics, Indian Institute of Technology Bombay, Powai, 
Mumbai 400 076, India}

\author{Sushant K. Raut}
\affiliation{
Department of Physics, Indian Institute of Technology Bombay, Powai, 
Mumbai 400 076, India}

\author{S. Uma Sankar}
\affiliation{
Department of Physics, Indian Institute of Technology Bombay, Powai, 
Mumbai 400 076, India}
\affiliation{Department of Theoretical Physics,
Tata Institute of Fundamental Research, 
Colaba, Mumbai 400 005, India}

\begin{abstract}
The recent discovery by the Daya-Bay and RENO experiments, 
that $\theta_{13}$ is nonzero and relatively large, 
significantly impacts existing experiments and the planning of 
future facilities. 
In many scenarios, the nonzero value of $\theta_{13}$ implies that 
$\theta_{23}$ is likely to be different from $\pi/4$. 
Additionally,
large detectors will 
be sensitive to matter effects on the oscillations of
atmospheric neutrinos, making 
it possible to determine the neutrino mass hierarchy and 
the octant of $\theta_{23}$.
We show that a $50$ kT magnetized 
liquid argon neutrino detector 
can ascertain the mass hierarchy with a significance larger than $4\sigma$ 
with moderate exposure times, and the octant
at the level of $2-3\sigma$ with greater exposure. 
\end{abstract}
\maketitle

\underline{\bf {Introduction:}} 
Neutrino oscillations in the standard three-flavor framework are 
parametrized by (a) two mass-squared differences 
$\Delta m^2_{j1}=m^2_j-m^2_1$, $j=2,3$; (b) three mixing angles
$\theta_{ij}$; and (c) one CP phase $\delta_{CP}$.  Over the past 
year, the T2K~\cite{t2k}, MINOS~\cite{minos}, and 
Double-Chooz~\cite{dc} experiments have provided evidence that the
mixing angle $\theta_{13}$ is nonzero and not significantly smaller 
than  the upper bound set by the Chooz~\cite{chooz} experiment. 
Recently, the Daya-Bay and RENO experiments 
have provided clinching evidence that
$\theta_{13} \neq 0$ at more than $5\sigma$: 
$\sin^2 2\theta_{13} = 0.089 \pm 0.011(stat) \pm 0.005(syst)$~\cite{db},
and
$\sin^2 2\theta_{13} = 0.113 \pm 0.013(stat) \pm 0.019(syst)$~\cite{reno}, 
respectively.

These results open new windows of opportunity for present-day 
or near-future neutrino experiments. If $\theta_{13}$ were tiny, 
an elaborate and expensive program would have been needed to 
measure the undetermined neutrino parameters. Its moderately 
large value allows for measurable
matter effects~\cite{matter} in both beam and atmospheric neutrino
experiments. This will yield better than 
anticipated  precision in measurements of neutrino parameters, and
impacts the planning of future neutrino facilities.
Specifically, it allows for a credible case to be made for 
building advanced technology, large-mass atmospheric neutrino 
detectors. The resolution of important outstanding questions, 
such as the {\it mass hierarchy} ({\it i.e.},
whether $m_3>m_1$ or $m_3<m_1$) and the {\it octant of $\theta_{23}$} 
({\it i.e.}, whether $\theta_{23}$ is larger or smaller than $\pi/4$), 
may come within the purview of such detectors.

The mass hierarchy plays a crucial role in the formulation 
of any theoretical effort designed to carry us beyond the 
standard model (SM) because models with a $\it normal$ 
hierarchy (NH) are significantly and qualitatively different 
from those with an $\it inverted$ hierarchy (IH); see, {\it e.g.}, 
\cite{albright}. Thus, knowledge of mass hierarchy can help in 
discriminating between classes of
models and consequently sharpen 
the focus of our search for new physics. A relatively large value of
$\theta_{13}$ suggests that  $\theta_{23}$ 
is unlikely to be maximal if the breaking of a $\mu-\tau$ 
exchange symmetry~\cite{lam} 
in the lepton sector causes $\theta_{13}$ to be nonzero.
A nonzero value of $\theta_{13}$ 
also opens the door to measurements of CP violation in the neutrino sector. 

Atmospheric neutrino detectors have been very important as 
discovery tools in the past. While they may not match the 
precision to pin down the energy and direction of an event  
characteristic of long baseline experiments, they have some 
strong compensating advantages. Atmospheric neutrinos offer 
a broad range in baselines ($L\sim 20$~km to 12500~km)  
and energy $E$ (100~MeV to 10~TeV) that can be tapped into 
by such an  experiment. An important consequence of this is 
the resolution or alleviation of degeneracies endemic to long 
baseline beam experiments~\cite{degen}.  Large-mass liquid argon 
time projection chambers (LATPCs) are based on one of the most 
promising technologies for charged particle detection. They have 
unprecedented capabilities for the detection of neutrino 
interactions, rare events, and dark matter because of their precise and 
sensitive spatial and calorimetric resolution. The 
ICARUS~\cite{liqar2} and ArgoNeuT~\cite{neut} detectors have 
provided excellent examples of the capabilities of liquid argon 
as a neutrino interaction detector medium.

In this Letter we study how well the hierarchy and octant 
may be determined by a liquid argon detector using atmospheric neutrinos 
in light of the recently reported nonzero value of 
$\theta_{13}$. 

\underline{\bf {Experimental specifications:}}
We consider a large liquid argon detector as discussed in 
\cite{Bueno:2007um,Rubbia:2004tz,Vissani}, which can detect 
charged particles with good resolution over the energy range 
of MeV to multi-GeV.  Magnetization over a 50-100 kT
volume has been proposed~\cite{Ereditato:2005yx} and we assume 
it in what follows. We also assume the following
detector resolutions over the GeV energy ranges relevant to our
calculations~\cite{Bueno:2007um,private}:
\bea
\sigma_{E_e} & = & \sigma_{E_{\mu}}  =  0.01\,, \nonumber \\   
\sigma_{E_{had}} & = & \sqrt{(0.15)^2/E_{had} + (0.03)^2}\,, \nonumber \\
\sigma_{\theta_e} & = & 0.03~{\rm{radians}} = 1.72^\circ, \nonumber \\
\sigma_{\theta_{\mu}} & = & \sigma_{\theta_{had}} = 0.04~{\rm{radians}} 
= 2.29^\circ\,, \nonumber  
\eea 
where $E_{had}$ is the hadron energy in GeV, $\sigma_E$ are the 
energy resolution widths in GeV and $\sigma_{\theta}$ are the angular 
resolution widths of electrons, muons, and hadrons.

Since  $E_{\nu} = E_{lep} + E_{had}$, 
the neutrino energy resolution width, for both $\nu_\mu$ and 
$\nu_e$, is  
\be
\sigma_{E_{\nu}} = {\sqrt{(0.01)^2 + (0.15)^2/(y E_{\nu}) + (0.03)^2}}\,,
\ee
where we have used $E_{had} = y E_{\nu}$, $y$ being the 
rapidity.
In our computation, we take the average rapidity in the GeV energy region 
to be 0.45 for neutrinos and 0.3 for antineutrinos \cite{Gandhi:1995tf}. 
The angular resolution of the detector for neutrinos can be worked out to 
be $\sigma_{\theta_{\nu e}} = 2.8^\circ$,
$\sigma_{\theta_{\nu\mu}} = 3.2^\circ$ \cite{private}. 
Charged lepton detection and separation ($e$ vs $\mu$) without 
charge identification is possible for $E_{lepton}>$ few MeV.
The charge identification capability of the detector is incorporated 
as discussed in \cite{GandhiLiqAr}. For electron events, we have 
conservatively assumed a 20$\%$ probability of  charge identification  
in the energy range $1 - 5$~GeV, and none for events with energies 
higher than 5 GeV. The muon charge identification capability of a 
LATPC is excellent for energies 1-10 GeV and we have 
assumed it to be $100\%$

The atmospheric fluxes are taken from the three-dimensional 
calculation in~\cite{Honda:2004yz}. The Earth matter profile 
defined in \cite{PREM} is used to take into account matter effects 
on the oscillation probabilities.

\underline{\bf {Hierarchy sensitivity:}}
The hierarchy sensitivity from atmospheric neutrinos is computed 
with  marginalization over the following test parameter ranges: 
\begin{itemize}
\item $\theta_{23}$ from $38^\circ - 52^\circ$ 
($\sin^2 \theta_{23} = 0.38 - 0.62$)\,,
\item $|\Delta m^2_{31}|$ from $(2.05 - 2.75) \times 10^{-3}$ eV$^2$\,, 
\item $\theta_{13}$ from $5.5^\circ - 11.0^\circ$ 
($\sin^2 2\theta_{13} = 0.04 - 0.14$)\,,
\item $\delta_{CP}$ from $0 - 2\pi$\,.
\end{itemize}
We take the true values to be 
$(|\Delta m^2_{31}|)_{tr} = 2.4 \times 10^{-3}$ eV$^2$, 
$(\delta_{CP})_{tr} = 0$, 
$(\sin^2 \theta_{23})_{tr} = 0.4, 0.5$, and $0.6$ which allows 
for both maximal and nonmaximal values, and $(\theta_{13})_{tr}$ 
over the $3\sigma$ range of recent measurements. 
The solar parameters $\Delta m^2_{21}$ and $\theta_{12}$ are fixed 
at the  values of $8 \times 10^{-5}$ eV$^2$ and $34^\circ$, 
respectively, since the effect of their variation within uncertainties 
is negligible.  The true value of $\delta_{CP}$ is chosen to be zero 
throughout these calculations, since the principal 
contribution to both the hierarchy and the octant sensitivity is from 
the muon survival probability, which has a weak dependence on
$(\delta_{CP})_{tr}$.
So the sensitivity is largely unaffected by the value of this 
parameter \cite{Gandhi}.  
We assume the true hierarchy to be normal and 
compute the ability
of the detector to rule out the inverted hierarchy and vice versa.
Flux uncertainties and systematic errors are included, using the 
method of pulls as described in \cite{Gandhi}:
flux normalization error 20$\%$, 
flux tilt factor, zenith angle dependence uncertainty 5$\%$, overall 
cross-section uncertainty 10$\%$, and overall systematic uncertainty 5$\%$. 
The number of bins is chosen to be 9 energy bins in the range 
$1-10$~GeV and 18 $\cos \theta_z$ bins in the range $-1.0$ to $-0.1$. 

\begin{figure*}[t]
\includegraphics[scale=0.7]{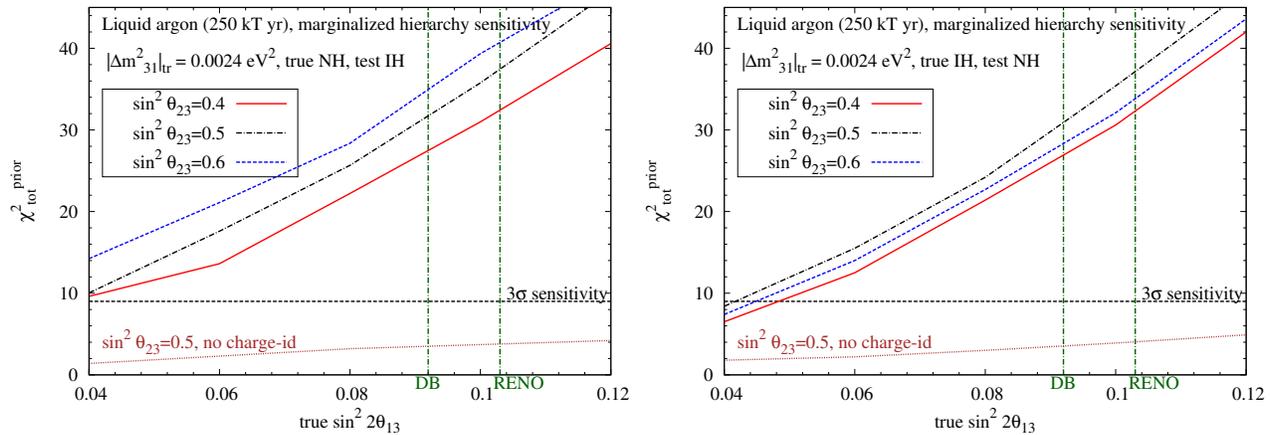}
\caption{{\em {Marginalized hierarchy sensitivity including priors in a liquid argon detector
with 250 kT yr exposure as a function of $(\sin^2 2\theta_{13})_{tr}$
for $(\sin^2 \theta_{23})_{tr} = 0.4, 0.5$, and $0.6$.
The vertical lines represent the Daya-Bay and RENO best-fit values.
The left panel is for a true normal hierarchy (NH) and the right
panel is for a true inverted hierarchy (IH).
Sensitivities for the nonmagnetized version of the detector 
for $\sin^2\theta_{23} = 0.5$ are also shown.
 }}}
\label{hier}
\end{figure*}

We also
take into account the uncertainties of neutrino parameters 
in the form of appropriate priors 
 with the following 1$\sigma$ ranges:  
$\sigma(|\Delta m^2_{31}|) = 0.05 |\Delta m^2_{31}|$,
$\sigma(\sin^2 2 \theta_{13}) = 0.01$, and 
$\sigma(\sin^2 2 \theta_{23}) = 0.02 \sin^2 2 \theta_{23}$.
The hierarchy sensitivity with priors ${\chi^2_{\mathrm tot}}^{\mathrm prior}$ is obtained 
from a combined minimization of $\chi^2_\mu+\chi^2_e + \chi^2_{\mathrm prior}$.

Figure~\ref{hier} depicts the hierarchy   
sensitivity for both NH and IH 
as a function of $(\theta_{13})_{tr}$ for
$(\sin^2 \theta_{23})_{tr}= 0.4$, $0.5$, and $0.6$. 
The Daya-Bay and RENO best-fit points are indicated in the figures.
In general, the hierarchy sensitivity from both muon and electron events
improves with an increase in the true value of $\theta_{13}$ and $\theta_{23}$. 
However, the sensitivity arising from muon events is  
considerably greater  due to 
the statistical advantage enjoyed by 
muon events and the superior charge identification capability for 
muons as compared to electrons \cite{Gandhi, petcov-schwetz}.
For both hierarchy and octant 
calculation below, sensitivity strongly depends on matter effects, 
the detection of which relies on charge identification. Additionally, 
as also discussed in \cite{petcov-schwetz, Gandhi, GandhiLiqAr}, 
the excellent angular 
resolution of  LATPC enables a full exploitation of the variations 
in the survival probability $P_{\mu\mu}$ (the main constituent of 
muon events) with baseline. 

In Table I we 
list the $\chi^2$ values for hierarchy sensitivity 
for $(\sin^2 2 \theta_{13})_{tr} = 0.1$ and
$(\sin^2 \theta_{23})_{tr} = 0.5$. 
To highlight the role of the measurement uncertainty in $\theta_{13}$,
we provide $\chi^2$ values for three cases: (a) no prior, (b) prior with 
$\sigma (\sin^2 2\theta_{13}) = 0.01$
(present uncertainty), and (c) $\sigma (\sin^2 2\theta_{13}) = 0.005$ 
(expected uncertainty in the near future). 
As expected, prior knowledge of
$\sin^2 2 \theta_{13}$ has a dramatic effect on the ability to 
determine the hierarchy.
However, an improvement in $\sigma (\sin^2 2\theta_{13})$ below the current
value does not give a concomitant improvement in the
sensitivity.
We note that without priors, the hierarchy sensitivity is much greater 
for a true NH, since in this case
the sensitivity arises from  resonant matter effects in the muon 
events, while for a true IH the
matter resonance occurs in the antimuon events. The atmospheric 
$\nu_\mu$ flux is about twice the $\bar{\nu}_\mu$ flux,
leading to greater sensitivity for NH. 
The prior terms tend to move the test parameters at the $\chi^2_{min}$ 
closer to their true values. This causes a tension between
$\chi^2_\mu+\chi^2_e$ and $\chi^2_{\mathrm prior}$ resulting in an overall increase 
in the total $\chi^2$.   
If the true hierarchy is inverted, then 
the test parameter values at the $\chi^2_{min}$ without priors  
are further removed from the true values as compared
to NH. Also, the sensitivity for $\sin^2 \theta_{23} =0.5$ becomes
superior to that for $\sin^2\theta_{23}$ = 0.6, as can be seen from the right panel of 
Fig.~\ref{hier};  we have checked that without the prior term the 
$\chi^2$s for IH for different values of $\theta_{23}$ 
follow the same behavior as that for NH. 
Clearly, the effect of priors is more significant for IH. 

For comparison we also present the sensitivity for 
a nonmagnetized version of the detector in Fig.~\ref{hier} 
for $\sin^2\theta_{23} = 0.5$. 
Note the crucial role played by the 
charge identification capability of a magnetized detector in discerning the 
mass hierarchy.

\begin{table}[t]
\begin{center}
\begin{tabular}{| c || c | c | }
\hline
{\sf {$\sigma(\sin^2 2\theta_{13})$}} & $\chi^2_{NH}$ & $\chi^2_{IH}$  \\
        \hline
        \hline
         No prior & 25.6  & 11.7  \\
            \hline
         0.01  & 35.8  & 35.4   \\
           \hline
         0.005 & 36.1  & 37.2   \\
            \hline
            \end{tabular}
            \caption[]{\footnotesize{
Values of marginalized hierarchy sensitivity (${{\chi^2}_{\mathrm tot}}^{\mathrm prior}$)
for various values of 
$\sigma(\sin^2 2 \theta_{13})$ with 250 kT yr exposure for 
$(\sin^2 2\theta_{13})_{tr} = 0.1$ and $(\sin^2 \theta_{23})_{tr} = 0.5$. 
The second column is for a true normal hierarchy and the third column is 
for a true inverted hierarchy.
 }}
            \end{center}
            \label{tab1}
            \end{table}

In summary, it is evident that if the value of
 $\theta_{13}$ lies within the $1\sigma$ range
preferred by Daya-Bay/RENO, then with a $50$~kT detector running for 5 years, 
an $ \sim 6\sigma$ hierarchy determination is 
possible for both hierarchies 
for $\sin^2 2\theta_{13} =0.1$
and $(\sin^2 \theta_{23})_{tr} = 0.5$
({\it cf.}~ Table I).  
Hence, even a reduced  
exposure of 100 kT yr may give 
a 4$\sigma$ hierarchy discrimination 
irrespective of whether the hierarchy is normal or inverted.

\underline{\bf {Octant sensitivity:}}
For this analysis, we adopt the viewpoint that once the hierarchy 
is determined with 250 kT yr exposure, the octant may be 
tackled with additional exposure
and a priori knowledge of the hierarchy.
We assume a total exposure of 
500 kT yr and consider the octant discrimination
separately for NH and IH. As with hierarchy discrimination, we  
include priors and perform a combined minimization of 
$\chi^2_\mu+\chi^2_e + \chi^2_{\mathrm prior}$.
The principal contribution to the octant sensitivity 
is again from muon events due to the $\sin^4 \theta_{23}$ term
in the  survival probability \cite{sandhya}. 

\begin{figure*}[t]
\includegraphics[scale=0.7]{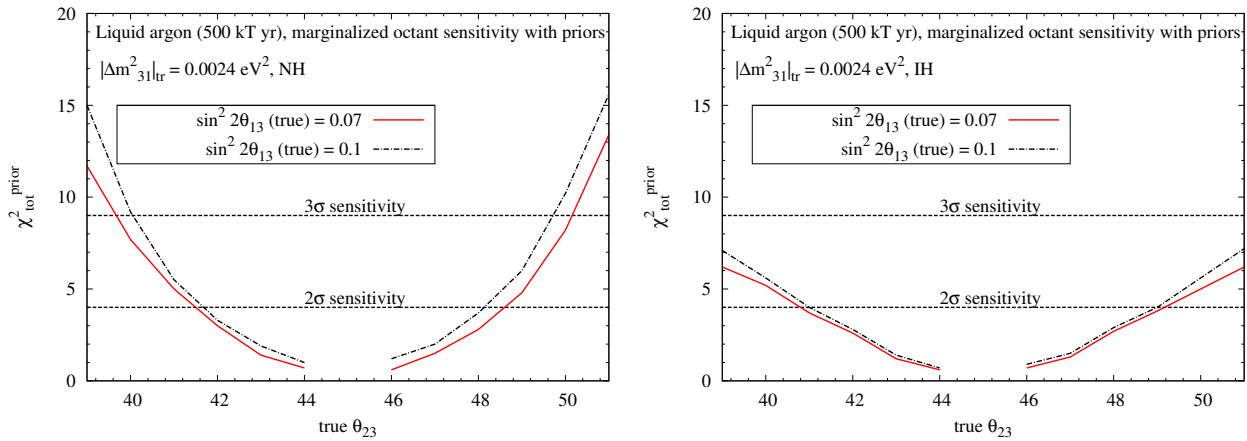}
\caption{{\em {Marginalized octant sensitivity including priors 
in a liquid argon detector
with 500 kT yr exposure as a function of true $\theta_{23}$
for $(\sin^2 2 \theta_{13})_{tr}
=0.07$ and $0.1$. The left panel is for NH and the right
panel is for IH.
 }}}
\label{octant}
\end{figure*}

In Fig. \ref{octant}, 
we plot the $\chi^2_{min}$ values  indicating
the ability of the experiment to rule out the wrong octant for 
two values of $\theta_{13}$ close to the Daya-Bay and RENO best fits.
In the left panel of Fig. \ref{octant}, the hierarchy is assumed to be normal 
and for the right panel it is inverted. 


We see that a $2\sigma$ discrimination is possible for 
$|\theta_{23} - \pi/4| > 3.5^\circ$ for values of $\theta_{13}$ 
close to the present best-fit, if the hierarchy is normal. That is, 
the octant discrimination is possible only if the value of 
$\sin^2 2 \theta_{23}$ is less than $0.985$. For
larger values of $\theta_{23}$, octant discrimination is  difficult.
For smaller 
values of $\sin^2 2 \theta_{23}$, the value of $\chi^2_{min}$
increases rapidly and a $3 \sigma$ octant discrimination 
is possible for $|\theta_{23} - \pi/4| > 5^\circ$ or
$\sin^2 2 \theta_{23} < 0.97$ for a normal hierarchy. 
If the hierarchy is inverted, octant sensitivity is worse, and only a $2\sigma$ discrimination is  possible for 
$|\theta_{23} - \pi/4| > 4^\circ$ or $\sin^2 2 \theta_{23}< 0.98$.


\underline{\bf {Conclusions:}}
The nature of the neutrino mass hierarchy and the octant of 
$\theta_{23}$ are vital to our efforts to build theories beyond 
the SM.  With regard to these two issues, we have explored the 
implications of the recent Daya-Bay and RENO  results on $\theta_{13}$  
for the planning of future experiments. In particular, we have 
demonstrated the exceptional
capability of a large-mass magnetized LATPC to determine 
the hierarchy to high significance with
moderate exposure times. 
The detector is also  sensitive to the octant 
of $\theta_{23}$ although with a lower significance. Our results highlight
the superior capability of a magnetized detector as compared to one
without magnetization.
 
{\it Acknowledgments:} This work was supported by 
U.S. DOE Grants No. DE-FG02-95ER40896 and DE-FG02-04ER41308,
and U.S. NSF Grant No. PHY-0544278. RG and PG acknowledge the 
support of the  XI Plan Neutrino Project under DAE. RG 
acknowledges the support of  the Indo-US Joint Centre on 
{\it Physics Beyond the Standard Model}-JC/23-2010. 
He is also grateful to the phenomenology 
group at the University of Wisconsin-Madison and the theory 
division at CERN for their hospitality while this work was in progress.
\vskip1cm


\end{document}